\documentclass[letter]{aa} 

\usepackage{graphicx}
\usepackage{color}
\usepackage{soul}
\usepackage{txfonts}
\usepackage{natbib,twoopt}
\usepackage[breaklinks=true]{hyperref} 
\bibpunct{(}{)}{;}{a}{}{,} 
\makeatletter
\newcommandtwoopt{\citeads}[3][][]{\href{http://adsabs.harvard.edu/abs/#3}%
{\def\hyper@linkstart##1##2{}%
\let\hyper@linkend\@empty\citealp[#1][#2]{#3}}}
\newcommandtwoopt{\citepads}[3][][]{\href{http://adsabs.harvard.edu/abs/#3}%
{\def\hyper@linkstart##1##2{}%
\let\hyper@linkend\@empty\citep[#1][#2]{#3}}}
\newcommandtwoopt{\citetads}[3][][]{\href{http://adsabs.harvard.edu/abs/#3}%
{\def\hyper@linkstart##1##2{}%
\let\hyper@linkend\@empty\citet[#1][#2]{#3}}}
\newcommandtwoopt{\citeyearads}[3][][]%
{\href{http://adsabs.harvard.edu/abs/#3}
{\def\hyper@linkstart##1##2{}%
\let\hyper@linkend\@empty\citeyear[#1][#2]{#3}}}
\makeatother

\begin{document}

   \title{Regaining the FORS: optical ground-based transmission spectroscopy of the exoplanet WASP-19b with VLT$+$FORS2}
	\titlerunning{Regaining the FORS}

   \author{E. Sedaghati\inst{1,2}
   			\and
			H.M.J. Boffin
          \inst{1}
          \and
          Sz. Csizmadia
          \inst{2}
          \and
          N. Gibson
          \inst{3}
			\and  \\
          P. Kabath
          \inst{4}
          \and
          M. Mallonn
          \inst{5}
                    \and
          M.E. Van den Ancker
          \inst{3}
                    }

   \institute{
   ESO, Alonso de C\'ordova 3107, Casilla 19001, Santiago, Chile\\
              \email{esedagha@eso.org; hboffin@eso.org}
         	\and
            Institut f\"ur Planetenforschung, Deutsches Zentrum f\"ur Luft- und Raumfahrt, Rutherfordstr. 2, 12489 Berlin, Germany
            \and
            ESO, Karl-Schwarzschild-str. 2, 85748 Garching, Germany
            \and
            Astronomical Institute ASCR, Fri\v{c}ova 298, Ond\v{r}ejov, Czech Republic
            \and
            Leibniz-Institut f\"ur Astrophysik Potsdam, An der Sternwarte 16, 14482 Potsdam, Germany
}

   \date{Received February 5, 2015; accepted March 12, 2015}

 \abstract{In the past few years, the study of exoplanets has evolved from being pure discovery, then being more exploratory in nature and finally becoming very quantitative. In particular, transmission spectroscopy now allows the study of exoplanetary atmospheres. Such studies rely heavily on space-based or large ground-based facilities, because one needs to perform time-resolved, high signal-to-noise spectroscopy. The very recent exchange of the prisms of the FORS2 atmospheric diffraction corrector on ESO's Very Large Telescope should allow us to reach higher data quality than was ever possible before. With FORS2, we have obtained the first optical ground-based transmission spectrum of WASP-19b, with 20 nm resolution in the 550--830 nm range. For this planet, the data set represents the highest resolution transmission spectrum obtained to date. We detect large deviations from planetary atmospheric models in the transmission spectrum redwards of 790 nm, indicating either additional sources of opacity not included in the current atmospheric models for WASP-19b or additional, unexplored sources of systematics. Nonetheless, this work shows the new potential of FORS2 for studying the atmospheres of exoplanets in greater detail than has been possible so far.}


   \maketitle
%

\section{Introduction}
Transiting exoplanets provide a wealth of information for studying planetary atmospheres in detail, particularly via spectroscopy. During a planetary transit, some of the stellar light passes through the limb of the planetary disc, where the presence of an atmosphere allows it to be indirectly inferred. When observed at different wavelengths, the transit depth, which is directly linked to the apparent planetary radius, may vary, providing constraints on the scale height of the atmosphere, the chemical composition, and the existence of cloud layers \citep{SS1998,SS2000,Brown2001,Burrows2014}. Such measurements require extremely precise relative photometry in as many wavebands as possible and as such can only be done using space telescopes or large ground-based facilities. 

The FOcal Reducer and low-dispersion Spectrograph (FORS2) attached to the 8.2-m Unit Telescope 1, is one of the workhorse instruments of ESO's Very Large Telescope \citep{1998Msngr..94....1A}. Using its capability to perform multi-object spectroscopy, \citet{Bean2010} show the potential of FORS2 in producing transmission spectra  for exoplanets even in the mini-Neptune and super-Earth regimes. They obtained the transmission spectrum of GJ1214b between wavelengths of 780 and 1000 nm, showing that the lack of features in this spectrum rules out cloud-free atmospheres composed primarily of hydrogen. Except for this pioneering result, all further attempts to use FORS2 for exoplanet transit studies have apparently failed, however, most likely becuase of the systematics introduced by the degradation of the antireflective coating of the prisms of the longitudinal atmospheric dispersion corrector \citep[LADC;][see also Moehler et al. 2010]{Berta2011}. A project was therefore started at ESO Paranal to make use of the available decommissioned twin instrument FORS1 \citep{BoffinMess2015}. The FORS2 LADC prisms were replaced by their FORS1 counterparts, which had their coating removed. This resulted in a transmission gain of 0.05 mag in the red to 0.1 mag in the blue, most likely because the uncoating of the LADC largely eliminates the contribution of scattered light from the previously damaged antireflective coating. As a further test of the improvement provided by the prism exchange, we also observed a transit of the exoplanet WASP-19b \citep{Hebb2010}. WASP-19 is a 12.3 magnitude G8V star, that hosts a hot Jupiter with a mass of 1.17 Jupiter masses (M$_{\rm J}$) and an orbital period of 0.789 days, making it the Jupiter-like planet with the shortest orbital period known and one of the most irradiated hot-Jupiters discovered to date. Owing to its short orbital period, and subsequently brief transit duration of $\sim$1h30, WASP-19b was an ideal target for assessing the impact of the prisms' exchange on the FORS2 performance.

\begin{figure}[tbp]
\includegraphics[width=0.95\linewidth]{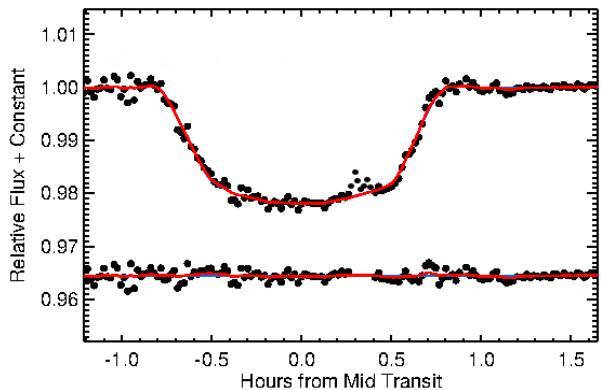}
\caption{Normalised, detrended broadband light curve of WASP-19 compared with the best fit obtained with the red noise model included (red line). The residuals of the light curve compared to the models are also shown. In the residual plot, the possible spot-affected data points are removed, because they are given zero weight.}
\label{fig:model}
\end{figure}

\section{Observations}
We observed WASP-19 between 2014 November 16 05:16 UT and 08:49 UT with FORS2, under thin cirrus, in multi-object spectroscopic mode (MXU), using a mask with slits 30\arcsec\ long and 10\arcsec\ wide placed on WASP-19, as well as on six reference stars. 
Data were binned ($2\times2$), yielding a scale of 0.25\arcsec\ per pixel. Grism 600RI (with the order sorter filter GG435) was used, leading to a wavelength coverage\footnote{The exact wavelength coverage depends on the position of the star on the CCD, so it varies from one star to the next. The values mentioned are the actual ones, which we use in our analysis.} of about 550 to 830 nm. Our observations covered a full transit, while the object moved from airmass 2.5 to 1.2, with the seeing varying between 0.8\arcsec\ and 2.2\arcsec. The high airmass at the start, in combination with the cirrus, is most likely the cause of the larger scatter in the light curve prior to ingress (Fig.~\ref{fig:model}). Using the MIT CCD, time series of the target and reference stars were obtained using 30s exposures until 08:18~UT\footnote{Beyond this point, the exposure time was reduced to 20s owing to the risk of saturation, but these data could not be used.}. The LADC was left in park position during the whole observing sequence, with the two prisms fixed at their minimal separation (30 mm). In total, we had 155 useful exposures, 85 of which were taken in transit. The typical signal-to-noise ratio of WASP-19 on one spectrum was about 300 at the central wavelength.

The data were reduced using IRAF, and one-dimensional spectra of the target and reference stars were extracted. Examples of such spectra are shown in the online Fig.~\ref{fig:spectra}. Stars in apertures 1 and 2 were not considered further in the analysis, since their wavelength coverages were either too red or too blue compared to WASP-19. For the remaining five comparison stars, we constructed differential white (or broadband, hence collapsing all the spectral channels into one) light curves of WASP-19 with respect to each of these, as well as all possible combinations, and measured the out-of-transit intrinsic scatter. It was found that using only the reference Star 7  provided the most accurate light curve, regardless of the detrending process employed, and this is what is then used in the rest of this paper. We conjecture that this is due to the low counts for Stars 4 and 5, and the significant colour difference with Star 6, which leads to differential atmospheric diffraction. Because our observations span a wide range of airmasses and there is clearly a difference in spectral type between our comparison star and WASP-19, we expect to see some smooth variation, owing to the colour term. This is then corrected via the atmospheric extinction function (from Bouguer's law), given by\begin{equation} \label{eq:extinction}
\text{F}_{obs} = \text{F}_0 \exp\Big[(k_1+k_2c)X\Big],
\end{equation}
 where $k_1$ and $k_2(\lambda)$ are the first- and second-order atmospheric extinction coefficients, respectively, $c$ is the stellar (B-V) colour index, and $X$ the airmass. This detrending process is shown for some of the spectrophotometric channels in the online Fig. \ref{fig:raw_LCs}, with the final broadband and spectroscopic light curves shown in Figs.~\ref{fig:model} and \ref{fig:spec LCs} (online), respectively. The post-egress, out-of-transit residuals in this light curve are 760 $\mu$mag. This value \citep[close to 579 $\mu$mag estimated from photon noise following the formalism of][]{gillon2006} and the fact that a single extinction function is sufficient to model, in large part, the correlated noise is a clear indication that the systematics that affected similar FORS2 observations in the past have been significantly reduced, so this instrument is now ready for detailed study of transiting exoplanets\footnote{Strictly speaking, however, other, possibly longer lasting transit observations are required to prove this point.}. This is what we do now.

\begin{table}[tp]
\centering
\caption{\label{tab:model}Planetary parameters derived from modelling the broadband light curve of the WASP-19b transit.}
\begin{tabular}{l  l}
\hline \hline
Parameter & Value\\
\hline
Scaled semi-major axis, $a/R_{\star}$ & 3.656 $^{+0.086}$$_{-0.097}$ \\
Scaled planetary radius, $R_p/R_{\star}$ & $0.1416 ^{+0.0019}$$_{-0.0018}$ \\ 
Inclination, \textit{i} & 80.36$^{\circ}$ $^{+0.76}$$_{-0.81}$ \\
Linear LD coefficient, $\gamma_1$ & 0.391 $^{+0.092}$$_{-0.096}$ \\
Quadratic LD coefficient, $\gamma_2$ & 0.225 $^{+0.052}$$_{-0.050}$\\
Mid-transit, $T_c~+2456977$ JD & 0.77722 $\pm$ 1.3$\times 10^{-4}$  \\
Uncorrelated (white) noise, $\sigma_w$ & 697 $\pm$ 65 $\mu$mag\\
\hline
\end{tabular}
\end{table}

\section{Results}
\subsection{Broadband light curve}

The normalised and detrended broadband light curve, obtained by integrating the spectra from 550 to 830 nm, was modelled using the {\tt Transit Analysis Package} (TAP) from \citet{gazak2011}, which is based on the formalism of \citet{MandelAgol2002}, as well as from \citet{caterwinn2009} for the wavelet-based treatment of the correlated (red) noise. The package is built on an MCMC code utilising a Metropolis-Hastings algorithm within a Gibbs sampler. The code has the ability to compute multiple MCMC chains in a single iteration and extend the chains until convergence is apparent. Eventually, a Gelman-Rubin statistic \citep{ford2006} was used to test for non-convergence by checking the likelihood that multiple chains have reached the same parameter space.  Once the state of convergence was reached, Bayesian inference is performed to quote the median solutions and the 1-$\sigma$ confidence levels. Parallel to this modelling process, we fitted the data again with another code, written by one of us (SzCs), using a combination of a genetic algorithm (performing a harmony search that finds a good local or a global minimum-like solution) and simulated annealing that refines this solution and performs the error estimation. The agreement in the planetary parameters is further testament to the reliability of our conclusions.

As part of our analysis, we tested the broadband and the individual spectrophotometric light curves for levels of correlated noise with the formalism described by \citet[][see online Fig. \ref{fig:red noise}]{pont2006}, where the maximum deviation from the white noise model is 65 $\mu$mag. Based on Eqs. 5 \& 6 of \citet{gillon2006}, we also estimated a red noise level of $\sigma_r$= 169 $\mu$mag. This prompted us to model and analyse this correlated noise using the wavelet decomposition method mentioned above, implemented in  TAP. For a complete description of wavelet mathematics applied to astronomical data see \citet{caterwinn2009}. Strictly speaking, however, this method can underestimate the level of correlated noise present in the data if the systematics are not time-correlated with a power spectra $\sim 1/f^\gamma$. Using a more general approach, such as a Gaussian process (GP) model introduced in \citet{gibson2012}, many alternative models of time-correlated noise can be explored, as can the effects of systematics that are not stationary in time, i.e. systematics that are functions of auxiliary parameters, such as the seeing and position of the star on the CCD. \citet{gibson2013} compare wavelet and Gaussian process (GP)  methods, showing that GPs provide more conservative estimates of the uncertainties even for purely time-correlated, stationary noise. However, these effects are only marginally present in our data (as shown in the online Figs. \ref{fig:red noise}). The modelling process is done with zero weight given to seven data points deemed to have been affected by the planet crossing a stellar spot (see \S \ref{sec:spot}). The results of our modelling are shown in Fig.~\ref{fig:model}, and the derived parameters are given in Table~\ref{tab:model}. In the modelling process, all the physical parameters are assumed to be free, apart from the period, eccentricity, and the argument of periapsis (fixed to 0.7888399 days, 0$^{\circ}$ and 90$^{\circ}$, respectively), which cannot be determined from our single transit observations.

The values we derive are in good agreement with what was found by others. As far as the fractional planetary radius is concerned, our value agrees with \citet{Mancini2013}, \citet{reed2013}, and \citet{Huitson2013}. We can thus be confident that our FORS2 observations are useful and not affected by systematics that cannot be corrected. Because WASP-19b is so close to its host star, orbiting at only 1.2 times the Roche tidal radius, the planet is most likely tidally deformed, and the radius we derived is likely to be underestimated by a few percentage points \citep{Lecomte11}.

\subsection{Transmission spectrum}

We have produced spectrophotometric light curves from 550 to  830 nm, with mostly 20 nm bandwidth at 10 nm intervals -- i.e. 27 light curves -- and modelled them in the same manner as before, shown in online Fig. \ref{fig:spec LCs}, while we fixed all the parameters that do not depend on wavelength to the white light curve solutions. As a result, only the relative planetary radii and the linear limb darkening coefficients are kept free. This process, also implies overlapping transmission spectrum points. Since the quadratic term is not well defined from the light curves, we assume the theoretical values, similar to that of the white light solution, as a Gaussian prior, allowing the parameter to vary under a Gaussian penalty error. The width of this distribution is taken as slightly larger than theoretically expected variations for this parameter within the given wavelength range. Furthermore, both correlated and uncorrelated noise components, along with a further airmass correction function (having two parameters), are also simultaneously modelled for all individual channels.

\begin{figure}
\centering
\includegraphics[width=0.95\linewidth]{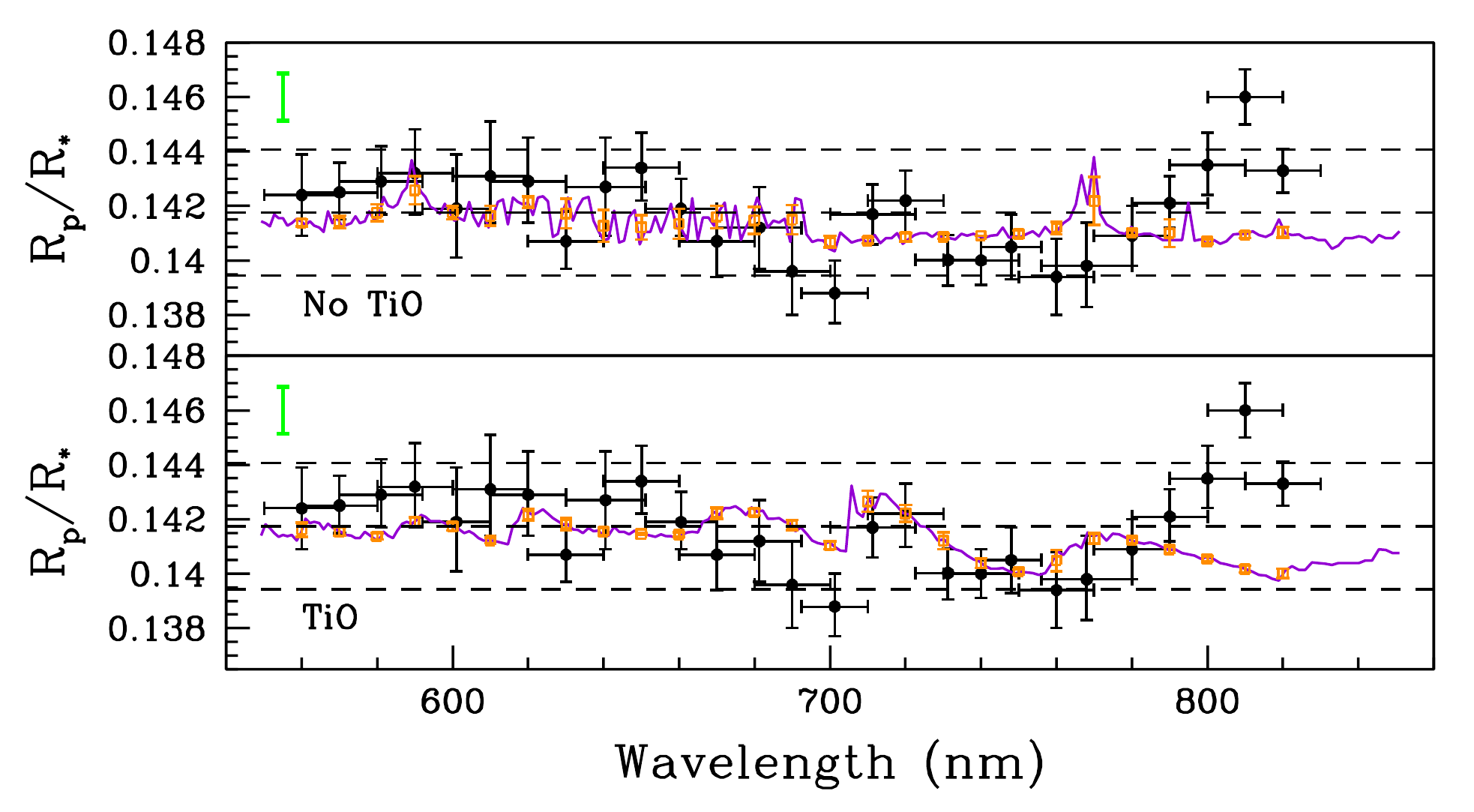}
\caption{Transmission spectrum of WASP-19b as measured with FORS2 (black dots, with error bars) compared to two models of planetary atmospheres, one with no TiO (top panel) and one with a solar abundance of TiO (bottom panel), from \citet{Burrows2010} and \citet{Howe2012}. We also estimated the mean value of the models in bin sizes of 20 nm (orange open squares). The dashed lines represent the weighted mean plus or minus three scale heights. Because of the overlapping spectral bins, only alternate pairs of transmission spectrum points contain unique information. The heavy, green bar on the left side shows the maximum error associated with unocculted spots.}
\label{fig:transmission2}
\end{figure}

The limb darkening coefficients were found to vary in a consistent manner as a function of wavelength when compared to theoretically calculated values \citep{claret2011}, shown in the online Fig. \ref{fig:linear LD}. The planetary radii obtained are shown as a function of wavelength in Figs.~\ref{fig:transmission2} and \ref{fig:transmission} and compared with previous values from the literature and to models of planetary atmospheres \citep{Burrows2010,Howe2012}.

The modelled uncorrelated noise is 900$\pm$50 $\mu$mag for all the spectroscopic channels. Since the noise in the data is predominantly white, the MCMC algorithm is not able to constrain the red noise, $\sigma_r$, for the various binned light curves, so we do not quote the determined value. The dominance of the uncorrelated noise is a testament to the minimal impact of the systematics upon our observations. It is also important to note that since the noise components, as well as the detrending functions, are modelled using the MCMC code, their uncertainties have been fully accounted for when quoting our final error budget estimations. Furthermore, the uncertainties in coefficient determination of the atmospheric extinction function used initially to correct the slow trend in the light curves have also been propagated into our error values. We are therefore confident that the relative error bars for our planetary radii values are realistic and comprehensive. The numerical results from the above modelling processes are shown in the online Table \ref{tab:spec results}.

From online Fig. \ref{fig:transmission}, we see that our measurements are those with the highest spectral resolution\footnote{It must be stressed that only alternate pairs of data points contain unique information, so we only claim to produce a transmission spectrum at a resolution of 20 nm.} obtained for WASP-19b in the optical. They clearly agree with the values determined with HST by \citet{Huitson2013}. In Fig.~\ref{fig:transmission2}, we compare our values with two models from \citet{Burrows2010} and \citet{Howe2012}, which were also used by \citet{Huitson2013}, after having assumed solar metallicity for chemical mixing ratios and opacities, as well as local chemical equilibrium.  Our measurements redwards of 790 nm do not seem to fit any of the models and deviate from them at more than 5-$\sigma$. Below 790 nm, our data is compatible, within 2-$\sigma$ with no variation in the planetary radius, and we can unfortunately not distinguish between the atmospheric models with and without TiO. HST observations from \citet{Huitson2013} favoured models lacking any TiO abundance, based on their low-resolution spectra covering the range 290-1030 nm. They rule out the presence of TiO features in the atmosphere, but additional WFC3 transmission spectroscopy in the infrared shows absorption feats owing to the presence of water. Similarly, \citet{Mancini2013} present ground-based, multi-colour, broadband photometric measurements of the transmission spectrum of WASP-19b, from which they determined the transmission spectrum over the 370-2350 nm wavelength range, concluding that there is no evidence for strong optical absorbers.

\subsection{Spotting a spot} \label{sec:spot}

Our broadband light curve, as well as our spectral light curves (Figs.~\ref{fig:model} \& \ref{fig:spec LCs}), shows a clear in--transit substructure, between 0.3 and 0.4 hours after mid-transit. This is possibly the signature of a spot (or group of spots) crossing by the transiting planet \citep{2003ApJ...585L.147S,2007A&A...476.1347P,Rabus09} -- such events have already been seen in the photometric light curves of WASP-19 by \citet{reed2013} and \citet{Mancini2013}. We used their code, {\tt PRISM+GEMC}, to model this spot and present the results in online Table~\ref{tab:spot} and Fig. \ref{fig:spot model}. The modelling with the spot provides very similar values for the planetary transit parameters, and we have checked that this modelling does not change our conclusions concerning the transmission spectrum. We note that the latitude of the spot we find is not very different from what was seen by \citet{reed2013} and \citet{Mancini2013}, although ours is slightly larger but shows a very different contrast: while \citet{reed2013} find a relatively bright spot with a contrast of 0.76--0.78, and \citet{Mancini2013} have values between 0.35 and 0.64, depending on the colour, we find a rather dark spot with fairly high contrast of 0.30.  
However, a word of caution is necessary here, since this feature could also be due to a sudden variation in seeing or systematics seen in all the reference stars at the same time. As a check, we used the average of the spot contrast values, modelled from the light curves, for the first and last four channels (blue to red) to estimate the spot temperature difference between the extremes of our transmission spectrum. The variation is significant at the 4-$\sigma$ level, although the actual change is four times smaller than found by \citet{Mancini2013} between the $r\arcmin$ and $i\arcmin$ bands, and we are thus left without any conclusion. We explored further avenues for causes of this feature, but the list of possibilities is by definition non-exhaustive owing to the nature of systematics. 

Besides the impact of spot occultation, there are other uncertainties introduced in the derived parameters, by possible presence of unocculted spots. Activity monitoring of the host star is required to quantify and rule out the impact of such events, before any significant claim about an atmospheric detection can be made. This, however, is an effect that is much more crucial to consider when dealing with multiple transits at large epoch separations. Since we are only dealing with relative radius variations, the only concern for us is the wavelength dependence of the spot contrast and its impact on the determinability of the out-of-transit baseline. Assuming a wavelength-dependent variation from blue to red of spot contrast and temperature ($\sim$ 100~K) similar to those found by \citet{Mancini2013} and an average spot size of $15^{\circ}$ (from this work and others), we estimate an upper limit uncertainty of 0.0017 in the determinability of the baseline (see Fig.~\ref{fig:transmission2}). This alone cannot explain the feature towards the red in the spectrum. The reader is referred to \citet{csizmadia2013} for further details on the unocculted spot impact. 

\section{Conclusions}
We have shown that following the prisms exchange of the LADC, FORS2 is now a competitive instrument for studying transiting planets using its multi-object spectroscopy mode. We observed one transit of WASP-19b as it passed in front of its host star and were able to model it. We thereby obtained the first ground-based, optical transmission spectrum of this planet, one of high resolution. Although the precision with which the planetary radii are determined needs to be improved by $\sim$ 50\% to distinguish between existing atmospheric models with and without TiO, we did detect clear significant variations in the deduced planetary radius at wavelengths redwards of 790 nm, which have so far not been explained by models. However, we still refrain from making any definitive conclusions here, because unexplored sources of systematics could also be responsible for these deviations. Finally, our observations possibly indicate the presence of a dark spot on WASP-19. 

Although we have limited ourselves to spectral bins of 20 nm wide, because we obtained our data under thin clouds, in a wide range of airmass and could finally only use one reference star, our analysis indicates that in good weather conditions and provided several reference stars are available, one can still improve upon the precision of the measurements and perhaps also  use smaller bins for transmission spectroscopy. Such data could allow distinguishing between competing planetary atmosphere models. In addition, these data could reveal additional information about the spectral signature of the atmospheric feature we are detecting longwards of 790 nm. We can thus only encourage readers to consider applying for FORS2 telescope time.

\begin{acknowledgements}
It is a pleasure to thank Guillaume Blanchard for his analysis of the feasibility of the prisms exchange and his precise and prompt work in uncoating the FORS1 LADC prims. Staff at Paranal were very efficient in making the exchange. We are very indebted to A. Burrows and J. Fortney for providing the results of their atmospheric models, and to C. Huitson for serving as an intermediary. We acknowledge the use of the publicly available {\tt PRISM+GEMC} and {\tt TAP} codes.  PK also acknowledges funding from M\v{S}MT \v{C}R, project LG14013, as well as Sz. Cs. the funding of the Hungarian OTKA Grant K113117. The observations of WASP-19b were obtained during technical time to check the exchanged prisms of the FORS2 LADC and are publicly available from the ESO Science Archive under Programme ID 60.A-9203(F).

\end{acknowledgements}

\Online
\appendix
\section{Additional tables and figures}

\begin{figure}[h]
\centering
\includegraphics[width=0.95\linewidth]{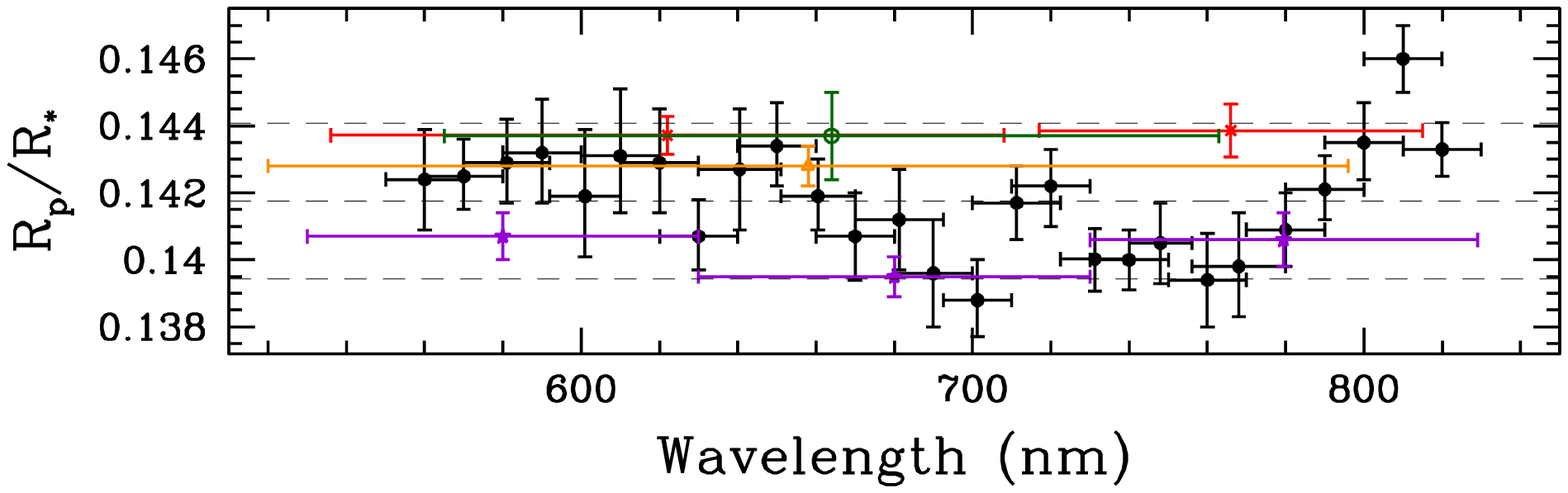}
\caption{Transmission spectrum of WASP-19b based on our FORS2 observations (black, filled dots), compared to values from \citet[][violet filled stars]{Huitson2013} (obtained spectroscopically), \citet[][red squares]{Mancini2013}, \citet[][green open circle]{Lendl2013},  and \citet[][orange open triangle]{reed2013} (from photometry). The vertical bars represent the errors in the fractional radius determination, while the horizontal bars are the FWHM of the passbands used. We note the high spectral resolution of the FORS2 data, compared to what was available until now. The dashed lines represent the weighted mean plus or minus three scale heights.}
\label{fig:transmission}
\end{figure}

\begin{figure}[h]
\includegraphics[width=9cm]{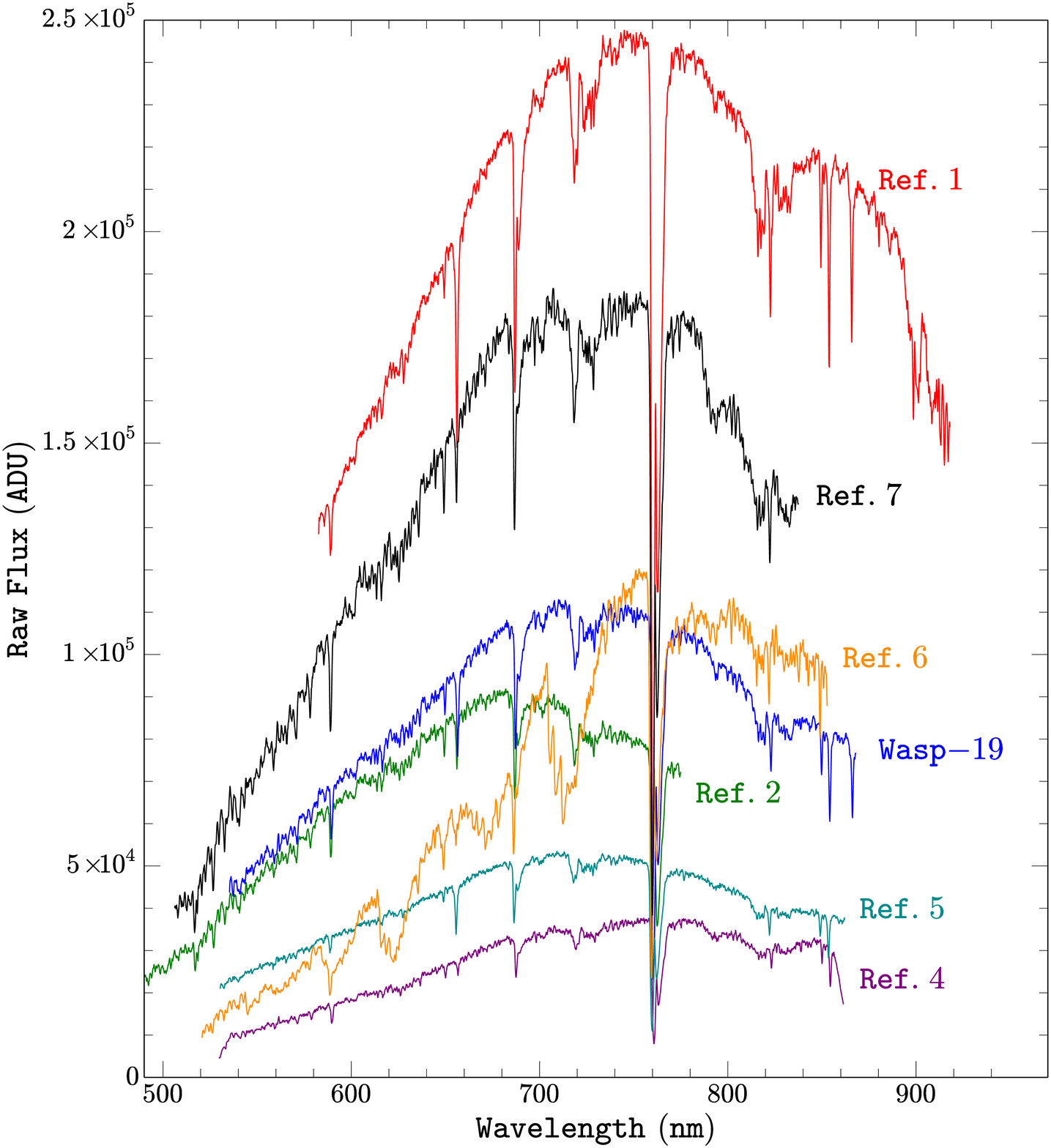}
\caption{\label{fig:spectra} Examples of the one-dimensional extracted spectra of the target and reference stars. The target star with the transiting planet, WASP-19, is shown in blue, while the target in aperture 7, which we use as comparison star, is shown in black.}
\end{figure}

\begin{figure}
\includegraphics[width=9cm]{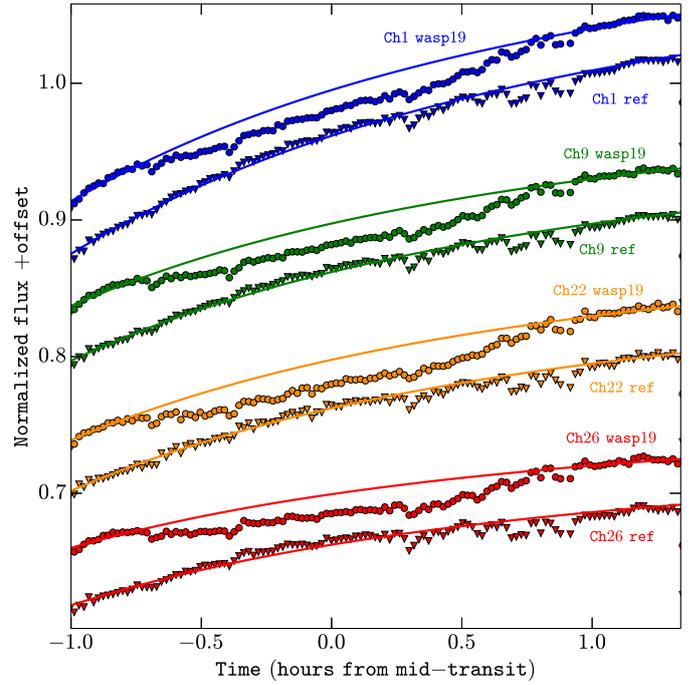}
\caption{\label{fig:raw_LCs} Raw light curves for four of the spectroscopic channels. The plots with circles are the light curves for WASP-19 and the triangles represent the reference star in aperture 7. The extinction functions (Eq. \ref{eq:extinction}), used for the purpose of detrending these raw light curves, are also shown as solid lines. An offset has been added to the light curves for clarity.}
\end{figure}

\begin{figure}
\centering
\includegraphics[width=\linewidth]{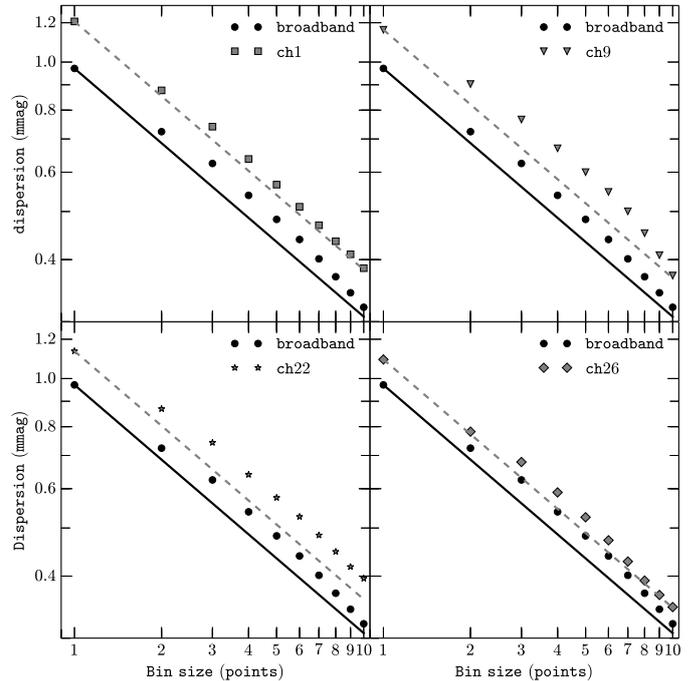}
\caption{Red noise impact, where standard deviation is calculated as a function of bin size, shown for the broadband and 4 spectrophotometric channels (separate panels) for sliding bins. Solid and dashed lines represent the $\sigma/\sqrt{n}$ relation, as expected from uncorrelated (white) noise. Deviation from this model, apparent as a straight line on the log-log plots, is evidence of minimal correlated (red) noise in our data. Using wavelets to model the red noise in all the channels ensured that the impact of this correlated noise is accounted for.}
\label{fig:red noise}
\end{figure}

\begin{figure}[ht]
\centering
	\includegraphics[width=\linewidth]{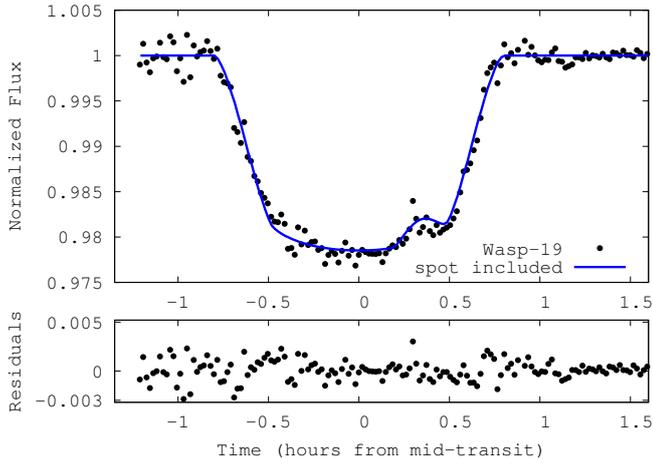}
	\caption{Broadband (white) light curve modelled with the {\tt GEMC+PRISM} code for the purpose of stellar spot characterization, values for which are shown in Table \ref{tab:spotmodel}.}
	\label{fig:spot model}
\end{figure}

\begin{figure*}[h]
\centering
\includegraphics[width=18cm]{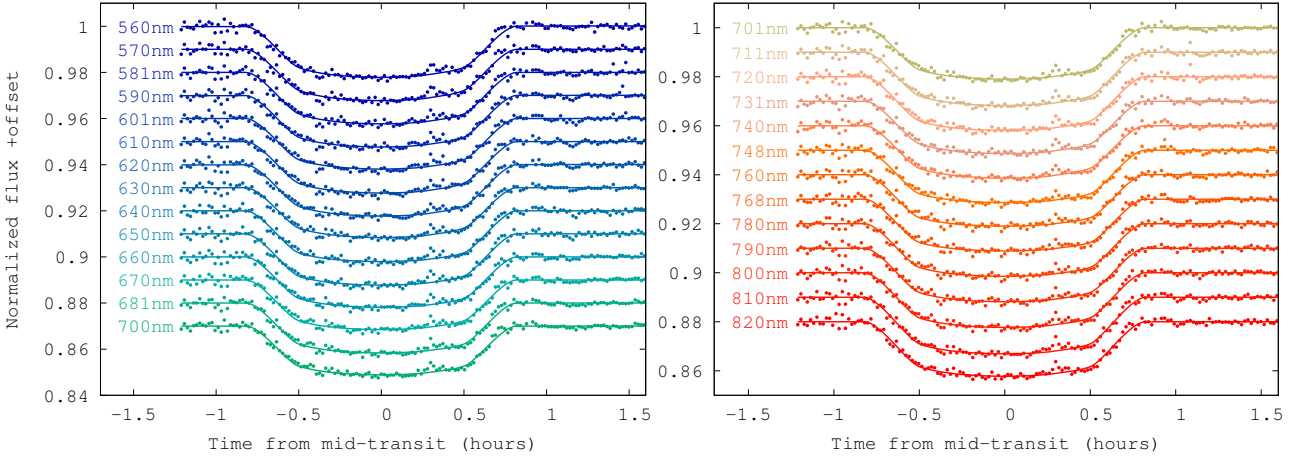}
\caption{Band-integrated spectrophotometric light curves for the transit of WASP-19b. The central wavelength for each channel is indicated on the left-hand side of each plot, where the integration width is mostly 20 nm. The modelled light curves for each channel are shown as solid lines.}
\label{fig:spec LCs}
\end{figure*}

\begin{figure}
\centering
	\includegraphics[width=\linewidth]{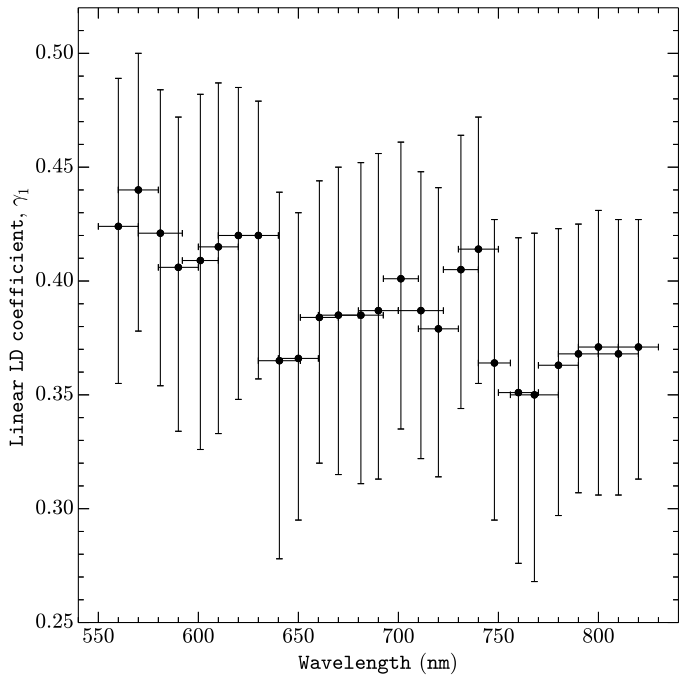}
	\caption{Variation in the linear limb darkening coefficient of the quadratic law, with wavelength, for the spectrophotometric channels. The values were allowed to vary between 0 and 1, as dictated by theory, for all the individual channels. The variations and the general trend agree with theoretical values calculated for photometric filters in our chosen range, from \citet{claret2011}.}
	\label{fig:linear LD}
\end{figure}

\begin{table*}[h]
\centering
\caption{All the determined planetary transit parameters for the modelling process of the spectrophotmetric channels for the transit of WASP-19b.  All other non-wavelength-dependent parameters are fixed to the broadband light curve model solution, values for which are given in Table \ref{tab:model}.}
\begin{tabular}{l c c c}
\hline \hline
Channel$^{a}$& R$_{p}$/R$_{\star}$ & Linear LD coefficient & Quadratic LD coefficient\\
(nm)& & $\gamma_1$ & $\gamma_2$ \\
\hline
\\
560	$\pm$ 10	   & 0.1424$^{+0.0015}$$_{-0.0015}$ 	& 0.424$^{+0.065}$$_{-0.069}$ 	& 0.228$^{+0.052}$$_{-0.053}$\\
570	$\pm$ 10	   & 0.1425$^{+0.0010}$$_{-0.0011}$ 	& 0.440$^{+0.060}$$_{-0.062}$ 	& 0.225$^{+0.053}$$_{-0.053}$\\
581	$\pm$ 11	   & 0.1429$^{+0.0012}$$_{-0.0013}$ 	& 0.421$^{+0.063}$$_{-0.067}$ 	& 0.230$^{+0.050}$$_{-0.052}$\\
590	$\pm$ 10	   & 0.1432$^{+0.0015}$$_{-0.0016}$ 	& 0.406$^{+0.066}$$_{-0.072}$ 	& 0.227$^{+0.052}$$_{-0.053}$\\
601	$\pm$ 9	           & 0.1419$^{+0.0018}$$_{-0.0020}$ 	& 0.409$^{+0.073}$$_{-0.083}$ 	& 0.221$^{+0.054}$$_{-0.051}$\\
610	$\pm$ 10	   & 0.1431$^{+0.0017}$$_{-0.0020}$ 	& 0.415$^{+0.072}$$_{-0.082}$ 	& 0.221$^{+0.053}$$_{-0.050}$\\
620	$\pm$ 10	   & 0.1429$^{+0.0015}$$_{-0.0016}$ 	& 0.420$^{+0.065}$$_{-0.072}$ 	& 0.221$^{+0.054}$$_{-0.052}$\\
630	$\pm$ 10	   & 0.1407$^{+0.0010}$$_{-0.0011}$ 	& 0.420$^{+0.059}$$_{-0.063}$ 	& 0.222$^{+0.052}$$_{-0.052}$\\
640.5	$\pm$ 10.5         & 0.1427$^{+0.0018}$$_{-0.0018}$ 	& 0.365$^{+0.074}$$_{-0.087}$ 	& 0.225$^{+0.052}$$_{-0.051}$\\
650	$\pm$ 10	   & 0.1434$^{+0.0012}$$_{-0.0013}$ 	& 0.366$^{+0.064}$$_{-0.071}$ 	& 0.225$^{+0.053}$$_{-0.052}$\\
660.5	$\pm$ 9.5          & 0.1419$^{+0.0010}$$_{-0.0011}$ 	& 0.384$^{+0.060}$$_{-0.064}$ 	& 0.222$^{+0.053}$$_{-0.052}$\\
670	$\pm$ 10	   & 0.1407$^{+0.0013}$$_{-0.0013}$ 	& 0.385$^{+0.065}$$_{-0.070}$ 	& 0.223$^{+0.054}$$_{-0.052}$\\
681.25	$\pm$ 11.2         & 0.1412$^{+0.0015}$$_{-0.0015}$ 	& 0.385$^{+0.067}$$_{-0.074}$ 	& 0.223$^{+0.055}$$_{-0.051}$\\
690	$\pm$ 10	   & 0.1396$^{+0.0016}$$_{-0.0016}$ 	& 0.387$^{+0.069}$$_{-0.074}$ 	& 0.225$^{+0.053}$$_{-0.053}$\\
701.25	$\pm$ 8.75         & 0.1388$^{+0.0011}$$_{-0.0012}$ 	& 0.401$^{+0.060}$$_{-0.066}$ 	& 0.222$^{+0.051}$$_{-0.051}$\\
711.25	$\pm$ 11.25        & 0.1417$^{+0.0011}$$_{-0.0011}$ 	& 0.387$^{+0.061}$$_{-0.065}$ 	& 0.221$^{+0.054}$$_{-0.051}$\\
720	$\pm$ 10	   & 0.1422$^{+0.0012}$$_{-0.0011}$ 	& 0.379$^{+0.062}$$_{-0.065}$ 	& 0.225$^{+0.052}$$_{-0.054}$\\
731.25	$\pm$ 8.75         & 0.1400$2^{+0.00095}$$_{-0.0009}$ 	& 0.405$^{+0.059}$$_{-0.061}$ 	& 0.229$^{+0.052}$$_{-0.053}$\\
740	$\pm$ 10	   & 0.1400$^{+0.0009}$$_{-0.0009}$ 	& 0.414$^{+0.058}$$_{-0.059}$ 	& 0.227$^{+0.053}$$_{-0.052}$\\
748	$\pm$ 8	           & 0.1405$^{+0.0012}$$_{-0.0012}$ 	& 0.364$^{+0.063}$$_{-0.069}$ 	& 0.222$^{+0.051}$$_{-0.051}$\\
760	$\pm$ 10	   & 0.1394$^{+0.0014}$$_{-0.0014}$ 	& 0.351$^{+0.068}$$_{-0.075}$ 	& 0.227$^{+0.051}$$_{-0.053}$\\
768	$\pm$ 12	   & 0.1398$^{+0.0015}$$_{-0.0016}$ 	& 0.350$^{+0.071}$$_{-0.082}$ 	& 0.224$^{+0.052}$$_{-0.052}$\\
780	$\pm$ 10	   & 0.1409$^{+0.0011}$$_{-0.0011}$ 	& 0.363$^{+0.060}$$_{-0.066}$ 	& 0.223$^{+0.051}$$_{-0.051}$\\
790	$\pm$ 10	   & 0.1421$^{+0.0009}$$_{-0.0010}$ 	& 0.368$^{+0.057}$$_{-0.061}$ 	& 0.224$^{+0.052}$$_{-0.052}$\\
800	$\pm$ 10	   & 0.1435$^{+0.0011}$$_{-0.0012}$ 	& 0.371$^{+0.060}$$_{-0.065}$ 	& 0.225$^{+0.051}$$_{-0.054}$\\
810	$\pm$ 10	   & 0.1460$^{+0.0010}$$_{-0.0010}$ 	& 0.368$^{+0.059}$$_{-0.062}$ 	& 0.229$^{+0.052}$$_{-0.053}$\\
820	$\pm$ 10	   & 0.1433$^{+0.0008}$$_{-0.0008}$ 	& 0.371$^{+0.056}$$_{-0.058}$ 	& 0.227$^{+0.052}$$_{-0.053}$\\
\hline
\multicolumn{4}{l}{$^{a}$~{\scriptsize The central wavelength and the width of some channels had to be adjusted slightly to avoid having integration limits that}}\\
\multicolumn{4}{l}{~~~{\scriptsize intersect with any telluric lines. Such an event could cause artificial trends in the light curve, due to uncertainties in the}}\\
\multicolumn{4}{l}{~~~{\scriptsize wavelength calibration.}}
\end{tabular}
\label{tab:spec results}
\end{table*}

\begin{table}[h]
\centering
\caption{\label{tab:spot}Planetary radius and the spot properties derived from modelling the broadband light curve of WASP-19b transit, where the occultation of a stellar spot is also included.}
\begin{tabular}{l  c  c }
\hline \hline
Parameter & Value\\
\hline
Scaled planetary radius, $R_p/R_{\star}$ & 0.1395 $\pm$ 0.0021  \\ 
Longitude of the spot centre (degrees) &	25.78 $\pm$	0.52 \\
Latitude of the spot centre (degrees) & 78.33 $\pm$	1.03 \\
Angular size of the Spot (degrees) & 20.6 $\pm$	0.4 \\
Spot contrast &	0.296 $\pm$	0.006 \\
\hline
\end{tabular}

\label{tab:spotmodel}
\end{table}


\begin{thebibliography}{}
\bibitem[Appenzeller et al.(1998)]{1998Msngr..94....1A} Appenzeller, I., Fricke, K., F{\"u}rtig, W., et al.\ 1998, The Messenger, 94, 1 

\bibitem[Bean et al.(2010)]{Bean2010} Bean, J.~L., Miller-Ricci Kempton, E., \& Homeier, D.\ 2010, \nat\ 468, 669

\bibitem[Bean et al.(2011)]{Bean2011} Bean, J.L., D{\'e}sert, J.-M., Kabath, P., et al.\ 2011, \apj\ 743, 92 

\bibitem[Berta et al.(2011)]{Berta2011} Berta, Z.K., Charbonneau, D., Bean, J., et al.\ 2011, \apj\ 736, 12 

\bibitem[Boffin et al.(2015)]{BoffinMess2015} Boffin, H.M.J., Blanchard, G., Sedaghati, E., et al. 2015, ESO Messenger

\bibitem[Brown(2001)]{Brown2001} Brown, T. 2001, \apj\ 553, 1006

\bibitem[Burrows (2014)]{Burrows2014} Burrows, A.S. 2014, \nat\ 513, 345

\bibitem[Burrows et al.(2010)]{Burrows2010} Burrows, A.S.,  Rauscher, E., Spiegel, D.S., Menou K. 2010, \apj\ 719, 341

\bibitem[Carter \& Winn(2009)]{caterwinn2009} Carter, J.A., Winn, J.N. 2009, \apj\ 704, 51

\bibitem[Claret \& Bloemen(2011)]{claret2011} Claret, A., Bloemen, S. 2011, \aap\ 529, A75

\bibitem[Csizmadia et al.(2013)]{csizmadia2013} Csizmadia, Sz., Pasternacki, T., Dreyer, C., Cabrera, J., Erikson, A., \& Rauer, H.\ 2013, \aap\ 549, A9

\bibitem[Ford (2006)]{ford2006} Ford, E.B. 2006, \apj\ 642, 505

\bibitem[Gazak et al.(2011)]{gazak2011}	Gazak, Z., Johnson, J., Tonry, J., Eastman, J., Mann, A., \& Agol, E. 2011, arXiv:1102.1036

\bibitem[Gibson et al.(2012)]{gibson2012} Gibson, N.P., Aigrain, S., Roberts, S., Evens, T.M., Osborne, M., Pont, F. 2012, \mnras\ 419, 2683

\bibitem[Gibson et al.(2013)]{gibson2013} Gibson, N.P., Aigrain, S., Barstow, J.K., Evans, T.M., Fletcher, L.N., Irwin, P.G.J. 2013, \mnras\ 428.3680G
	
\bibitem[Gillon et al.(2006)]{gillon2006} Gillon, M., Pont, F., Moutou, C., Bouchy, F., Courbin, F., Sohy, S., \& Magain P.\ 2006, \aap\ 459, 249

\bibitem[Hebb et al.(2010)]{Hebb2010} Hebb, L., Collier-Cameron, A., Triaud, A.~H.~M.~J., et al.\ 2010, \apj\ 708, 224 

\bibitem[Huitson et al.(2013)]{Huitson2013} Huitson, C.M., Sing. D.K., Pont, F., et al. 2013, \mnras\ 434, 3252

\bibitem[Howe \& Burrows(2012)]{Howe2012} Howe, A.R. \& Burrows, A.S. 2012, \apj\ 756, 176

\bibitem[Leconte, Lai \& Chabrier(2011)]{Lecomte11} Leconte, J., Lai, D. \& Chabrier, G. 2011, \aap\ 528, A41

\bibitem[Lendl et al.(2013)]{Lendl2013} Lendl, M., Gillon, M., Queloz, D., Alonso, R., Fumel, A., Jehin, E., \& Naef, D. 2013, \aap\ 552, A2

\bibitem[Mancini et al.(2013)]{Mancini2013} Mancini, L., Ciceri, S., Chen, G. 2013, \mnras\ 436, 2

\bibitem[Mandel \& Agol(2002)]{MandelAgol2002} Mandel, K., \& Agol, E.\ 2002, \apjl\ 580, L171 

\bibitem[Moehler et al.(2010)]{Moehler2010} Moehler, S., Freudling, W., M{\o}ller, P., et al.\ 2010, \pasp\ 122, 93 

\bibitem[Pont et al.(2006)]{pont2006} Pont, F., Zucker, S., \& Queloz, D. 2006, \mnras\ 373 (1), 231-242

\bibitem[Pont et al.(2007)]{2007A&A...476.1347P} Pont, F., Gilliland, R.~L., Moutou, C., et al.\ 2007, \aap\ 476, 1347

\bibitem[Rabus et al.(2009)]{Rabus09} Rabus, M., Alonso, R., Belmonte, J.~A., et al.\ 2009, \aap\ 494, 391 

\bibitem[Seager \& Sasselov(1998)]{SS1998} Seager, S., \& Sasselov, D.D. 1998, \apjl\ 502, 157

\bibitem[Seager \& Sasselov(2000)]{SS2000} Seager, S., \& Sasselov, D.D. 2000, \apj 537, 916

\bibitem[Silva(2003)]{2003ApJ...585L.147S} Silva, A.~V.~R.\ 2003, \apjl\ 585, L147 

\bibitem[Tregloan-Reed et al.(2013)]{reed2013} Tregloan-Reed, J., Southworth, J., \& Tappert, C.\ 2013, \mnras\ 428, 3671 
 
  \end{thebibliography}
\end{document}